\documentclass{ws-procs9x6}
\usepackage{bbding}
\usepackage{pifont}
\usepackage{amsmath}
\usepackage{amsfonts}
\usepackage{amssymb}
\usepackage{mathrsfs}

\def\bc{\begin{center}}
\def\nno{\nonumber}
\def\ec{\end{center}}
\def\be{\begin{eqnarray}}
\def\ee{\end{eqnarray}}

\def\dS{dS}

\def\R{I\!\!R}

\def\N{N \hspace{-0.7em}_{_{\sim}} ~}
\def\al{\alpha}

\def\dl{\delta}
\def\eps{\epsilon}
\def\ka{\kappa}
\def\la{\lambda}
\def\th{\theta}
\def\si{\sigma}

\def\Om{\Omega}
\def\del{\nabla}
\def\d#1#2{\frac{\displaystyle #1}{\displaystyle #2}}
\def\r{\partial}
\def\P{{\bf P}}
\def\K{{\bf K}}
\def\H{{H}}
\def\J{{\bf J}}
\def\N{{\bf N}}
\def\M{{\it M}}
\def\R{{\it R}}

\newcommand{\vect}[1]{\mbox{\boldmath $#1$}}

\newcommand\btd{\raise 2pt
\hbox{$\hat\bigtriangledown$}\hskip 1.5pt}
\newcommand\bt{\raise 2pt
\hbox{$\bigtriangledown$}\hskip 1.5pt}

\newcommand{\omits}[1]{}

\newcommand{\AdS}{${ A}d S$}

\begin{document}

\title{PRINCIPLE OF RELATIVITY, 24 POSSIBLE KINEMATICAL ALGEBRAS
AND NEW GEOMETRIES WITH POINCAR\'E SYMMETRY}

\author{Chao-Guang Huang$^*$}

\address{Institute of High Energy Physics,  Chinese Academy of
Sciences, Beijing 100049, and\\
Theoretical Physics Center for Science Facilities, Chinese Academy of
Sciences, Beijing 100049\\
$^*$E-mail: hangcg@ihep.ac.cn}

\begin{abstract}
From the principle of relativity with two universal invariant parameters $c$ and $l$,
24 possible kinematical (including geometrical and static) algebras can
be obtained. Each algebra is of 10 dimensional, generating the symmetry of a
4 dimensional homogeneous space-time or a pure space.  In addition to the
ordinary Poincar\'e algebra, there is another Poincar\'e algebra among the
24 algebras.  New 4d geometries with the new Poincar\'e symmetry are
presented.  The motion
of free particles on one of the new space-times is discussed.
\end{abstract}


\bodymatter

\section{Introduction}\label{aba:sec1}

The principle of relativity, laws of non-gravitational physics having the
same form in all inertial frames, is valid not only in the Minkowski
space-time but also in the de Sitter (dS)
space-time \cite{Lu-Guo, dSSR}.  Based on the principle of relativity and the postulate of the two
universal invariant parameters $(c,l)$, dS invariant special
relativity can be established in a dS space-time \cite{Lu-Guo, dSSR}, where $c$ is the vacuum speed
of light at the origin and $l$ is the dS radius.  For brevity, the principle of
relativity and the postulate of the two universal invariant parameters are known
as the principle of relativity with two universal invariant parameters, denoted
by $PoR_{c,l}$ \cite{GWZ,GHWZ}.

Very recently, in the study of the principle of relativity
with two invariant parameters, we construct 24 kinematical algebras,
including purely geometrical ones and static one \cite{GHWZ}.  Each algebra is of 10
dimensional.  The 11 of them are the algebras in
Bacry-L\'evy-Leblond theorem \cite{BLL}.  They are (Anti-)dS (${\frak d}_\pm$),
Poincar\'e ($\frak p$), (Anti-)Newton-Hooke ($\frak{n}_\pm$), Inhomogeneous SO(4) and
para-Poincar\'e (${\frak p'}_\pm$), Galilei ($\frak g$), Carroll ($\frak c$),
para-Galilei (${\frak g}'$) and static ($\frak s$) algebras.  The 3 of them correspond
to the Euclid geometry ($\frak{e}$), Riemann geometry ($\frak{r}$), and
Lobachevski geometry ($\frak l$), which can be obtained by relaxing the third assumption in
Bacry-L\'evy-Leblond theorem.  It is remarkable that
among the 10 new kinematical or purely geometrical algebras, there is another
Poincar\'e algebra\footnote{An algebra is said to be Poincar\'e one
if (1) it is isomorphic to $\frak{iso}(1,3)$ algebra, (2) the unique Abelean
ideal of the $\frak{iso}(1,3)$ algebra
is regarded as a translation sub-algebra and is divided into the time
translation and space translations as a 1d and a 3d representation,
respectively, of $\mathfrak{so}(3)$ sub-algebra of the
$\frak{so}(1,3)$ sub-algebra and (3) the algebra is invariant under the
suitably defined parity and time-reversal operation \cite{BLL}.}.

It is well known that the Poincar\'e symmetry is the foundation of Einstein's
special relativity, relativistic field theories in Minkowski space-time,
particle physics, as well as the Poincar\'e gauge theories of gravity, etc.
Conventionally, only the Minkowski space-time is invariant under
global Poincar\'e transformations.  It is natural to ask: what is the role
played by the new Poincar\'e symmetry.

The aim of the present talk is twofold.  One is to exhibit 24 kinematical algebras,
including purely geometrical ones and static one.  The other is to first present
the nontrivial 4d geometries which are invariant under the new Poincar\'e
transformations.  The structure of the new nontrivial 4d geometries will be explored
briefly.  The motion of free particles on the one of the new space-times is also
discussed.

The talk is divided into 6 parts.  After the introduction, I shall review
the inertial motion and Umov-Weyl-Fock-Hua (UWFH) transformations, show all
possible kinematical and geometrical algebras, present the new nontrivial
geometries, and study the motion of free particles, successively.  Finally, I
shall end my talk with the summary.


\section{Inertial Motions and UWFH transformations}

In a Cartesian coordinate system in a flat space-time (no matter whether it is relativistic
one or not), the motions satisfying
\be \label{urm}
\begin{cases}
x^i=x^i_0+v^i(t-t_0), &\\
v^i=\d {dx^i}{dt}={\rm consts.} & i=1,2,3
\end{cases}
\ee
or
\be \label{a0}
\d {d^2x^i}{dt^2}=0
\ee
are called the uniform rectilinear motions or inertial motions.
The set of observers in the space-time, moving according to Eq.(\ref{urm}) or
Eq. (\ref{a0}), make up an inertial frame, denoted by ${\cal F}$.  In a given flat
space-time, the forms of Eq.(\ref{urm}) and Eq.(\ref{a0}) are unchanged under
the linear coordinate transformation with 10 parameters.

It has been shown that the forms of Eq.(\ref{urm}) and Eq.(\ref{a0}) are unchanged
under the linear fractional transformations with a common denominator in the Beltrami
coordinate system in a(n) (A)dS space-time \cite{Lu-Guo,dSSR}.  In other words,
the above concept of inertial motions and inertial frame can be
generalized to the (A)dS space-times \cite{Lu-Guo,dSSR}.
Then, the principle of relativity can be generalized to the two
space-times.  Obviously, in the (A)dS space-time, the (A)dS radius $l$ is an
invariant parameter
in addition to the invariant speed of light $c$.  Therefore, the postulate of
invariant speed of light $c$ in Einstein's special relativity should be replaced
by the postulate of two invariant parameters.  The principle of relativity
and the postulate of an invariant parameter should be replaced by the principle
of relativity with two universal invariant parameters, $PoR_{c,l}$.

Furthermore, the forms of Eq.(\ref{urm}) and Eq.(\ref{a0}) are also unchanged under
the linear fractional transformations with a common denominator
in (Anti-)Newton-Hooke ((A)NH) space-times \cite{NH}.
Then, a simple question appears: what is the most general transformation ${\cal T}$ s.t.
\be
{\cal T}:\quad {x'}^\mu = f^\mu(x), x^0=ct, \mu=0,\cdots, 3,
\ee
preserving the form of Eq.(\ref{urm}) and Eq.(\ref{a0})?

The following theorem answers the
question.
\begin{theorem}
{\it The most general transformations are the linear fractional transformations:
\begin{equation}
{\cal T}:\quad  l^{-1}x'^\mu = \frac{A^\mu_{\ \nu} l^{-1} x^\nu +
a^\mu}{b^t  l^{-1}x + d} \label{eq:LFT}
\end{equation}
and
\begin{equation}
  det\ {\cal T}=\left| \begin{array}{rrcrr}
    A    & a \\
    b^t
     & d
  \end{array} \right|
  = 1,
\label{eq:det}
\end{equation}
 where $A=\{A^\mu _{~\nu}\}$  a $4\times 4$ matrix, $a,b$  $4\times 1$
 matrixes, $d\in R$
and $b^t=\eta b $ with $\eta_{\mu\nu}={\rm diag}(1,-1,-1,-1)$.}
\end{theorem}
The proof of the theorem can be found in Ref. \refcite{Fock,Hua,CQG}.  The question was
first raised and answered by Umov and Weyl\cite{Umov-Weyl}.  Fock and Hua also
studied the question in details in their books \cite{Fock,Hua}.  Therefore, we name
the transformations by Umov-Weyl-Fock-Hua (UWFH) transformations.

Clearly, all UWFH transformations form a group.  The number of generators of the
group is 24.
\omits{which is greater than the number of generators of Poincar\'e group.}
It implies that more possible space-times admit Eq.(\ref{urm})
and Eq.(\ref{a0}), as expected.

\section{Possible Kinematical Groups}
In order to clarify how many space-times admit Eq.(\ref{urm})
and Eq.(\ref{a0}), we begin with the Beltrami model of (A)dS spacetime
\be
ds_\pm^2=\left (\d {\eta_{\mu\nu}}{\si_\pm(x)}\pm\d {x_\mu x_\nu}
{l^2\si_\pm^2(x)}\right )dx^\mu dx^\nu,
\ee
where $x_\mu =\eta_{\mu\la} x^\la$ and
\be
\si_\pm(x)=\si_\pm(x,x) = 1 \mp l^{-2} x^\mu x_\mu >0.
\ee
In the above equations, the upper sign corresponds to dS space-time
and the lower sign to AdS space-time.  The (A)dS space-time is invariant
under the (A)dS transformations,
respectively.  The generators of (A)dS group are
\be
\begin{array}{l}{\bf P}^\pm_\mu =(\dl^\nu_\mu \mp l^{-2}x_\mu x^\nu)
\r_\nu ,\\
{\bf L}_{\mu\nu}=x_\mu {\bf P}_\nu - x_\nu {\bf P}_\mu=
x_\mu\r_\nu-x_\nu \r_\mu \in \frak{so}(1,3),\end{array}
\ee
or
\be\label{dS-gen3}
\begin{array}
{ll}H^\pm=\r_t\mp \nu^2 t x^\si \r_\si, &
{\bf P}^\pm_i=\r_i\mp l^{-2}x_ix^\si\r_\si, \quad \nu= c/l, \\
{\bf K}_i=t\r_i-c^{-2}x_i\r_t, &
{\bf J}_i=\frac 1 2 \eps_i^{\ jk}L_{jk}=\frac 1 2 \eps_i^{\ jk}(x_j\r_k-x_k\r_j),
\end{array}
\ee
where $H^\pm$ are called the Beltrami-time-translation generators,
${\bf P}^\pm_i$ are known as the Beltrami-space-translation generators,
${\bf K}_i$ and ${\bf J}_i$ are the Lorentz boost generators
and the space-rotation generators, as usual.  $H^\pm$ are
the scalar representation of the $\frak{so}(3)$ spanned by ${\bf J}_i$.
${\bf P}^\pm_i$ and ${\bf K}_i$ are vector representations of the $\frak{so}(3)$.

Now, we can write down the 24 generators for the group which
keeps Eq.(\ref{urm}) and Eq.(\ref{a0}).  They are \cite{GHWZ}: 4 kinds of
generators for time translation
\be
H^{\pm}= \r_t \mp \nu^{-2} t x^\mu\r_\mu, \qquad  H:=\r_t, \qquad H'=-\nu^{-2} t x^\mu\r_\mu ;
\ee
4 kinds of generators for space translation
\be
{\bf P}_i^{\pm}= \r_i \mp l^{-2} x_i
x^\mu\r_\mu,\qquad {\bf P}_i:=\r_i,
\qquad {\bf P}_i'=-l^{-2} x_i x^\mu\r_\mu ;
\ee
4 kinds of generators for boost
\be
\begin{array}{ll}{\bf K}_i:=t\r_i-c^{-2}x_i\r_t,  &\qquad
 {\bf K}_i^{\frak g}= t\r_i, \\
 {\bf K}_i^{\frak c}
= -c^{-2}x_i \r_t,& \qquad
{\bf N}_i= t\r_i+c^{-2}x_i\r_t;\end{array}
\ee
3 generators of rotation ${\bf J}_i$ as shown in Eq.(\ref{dS-gen3}) and
\be
\begin{array}{l}
R_{ij}=R_{ji}=x_i\r_j+x_j\r_i,  \qquad  (i < j) \\
M_0=t\r_t, \quad M_1=x^1\r_1, \quad M_2=x^2\r_2, \quad M_3=x^3\r_3.
\end{array}
\ee
Among them, two time-translation generators, two sets of space-translation generators and
two sets of boost generators are independent, respectively.  $H=\frac 1 2 (H^++H^-)$ and
${\bf P}_i=\frac 1 2 ({\bf P}^+_i+{\bf P}^-_i)$ are ordinary time- and space-translation
generators, respectively.  $H'=\frac 1 2 (H^+-H^-)$ and
${\bf P}'_i=\frac 1 2 ({\bf P}^+_i-{\bf P}^-_i)$ are known as pseudo-time- and
pseudo-space-translation generators, respectively \cite{GWZ,GHWZ}.
${\bf K}_i^{\frak g}$ are the Galilei-boost generators.
${\bf K}_i^{\frak c}={\bf K}_i - {\bf K}_i^{\frak g} $ are the Carroll-boost generators,
${\bf N}_i=2 {\bf K}_i^{\frak g}-{\bf K}_i$ are the geometrical-boost generators
\cite{GWZ,GHWZ}.  The set $\{T\} := (H^\pm, \P^\pm_i, \J_i,\K_i,\N_i,
M_0,M_i,R_{ij})$ spans a closed algebra,
\begin{eqnarray}
\begin{array}{l}
\, {[}\P^+_i, \P^-_j{]} =(1-\delta_{(i)(j)})l^{-2} R_{(i)(j)} -
2l^{-2}
\delta_{i(j)} (\M_{(j)}+\Sigma_\ka M_\ka), \\
\begin{array}{ll}
{[}\P^\pm_i, \M_j{]}= \delta_{i(j)}\P^\mp_{(j)},&
[\P^{\pm}_i, \R_{jk}]=-\delta_{ij}\P^{\mp}_k-\delta_{ik}\P^{\mp}_j,\\
{[}\H^+, \H^-{]} =  2{ \nu^2} \left(\M_0 + \Sigma_\ka \M_\ka\right),
&
[\H^\pm, \M_0]= \H^\mp,  \\
{[}\K_i, \M_0{]}=-\N_i, &   [\K_i, \M_j]= \delta_{i(j)} \N_{(j)},\\
{[}\K_i, \R_{jk}{]}= -\delta_{ij} \N_k-\delta_{ik}\N_j,& [\N_i, \M_0]= -\K_i,\\
{[}\N_i, \M_j{]}= \delta_{i(j)}\K_{(j)},&
{[}\N_i,\R_{jk}{]}=-\delta_{ij} \K_k-\delta_{ik} \K_j, \\
{[}\K_i,\, \N_j{]} =(\delta_{(i)(j)}-1)  c^{-2} R_{(i)(j)} -&
2\delta_{(i)j}c^{-2}\left(M_{0} - M_{(i)}\right),\\
{[}L_{ij}, \M_k{]}=\delta_{j(k)}  R_{i(k)}-\delta_{i(k)}R_{j(k)} ,&
{[}R_{ij}, \M_k{]}=\delta_{i(k)}L_{j(k)} + \delta_{j(k)}L_{i(k)},
\end{array}\\
\,
{[}L_{ij},\R_{kl}{]}=2(\delta_{ik}\delta_{jl}+\delta_{il}\delta_{jk})\,
(\M_i  - \M_j) + \delta_{ik}\R_{jl} +\delta_{il}\R_{jk}\\
\qquad \qquad \quad \ -\delta_{jk} R_{il}- \delta_{jl}R_{ik}, \\
\, {[}\R_{ij},R_{kl}{]} = -
\delta_{ik}L_{jl}-\delta_{il}L_{jk}-\delta_{jk}L_{il}-\delta_{jl}L_{ik},\\
\mbox{dS, AdS, Riemann, and Lobachevski algebraic relations,}
\end{array} \label{im4}
\end{eqnarray}
where
no summation is taken for the repeated  indexes in brackets.

It has been shown that there are 24 possible kinematical (including geometrical) algebras
\cite{GHWZ}, in which $\J_i$ serve as the space-rotation generators.  They include 4 pure
geometrical algebras and 1 static algebra.  The algebras, the sets of generators,
and the commutators are listed in Table 1.  In Table 1, ${\cal H}$, $\P$,
$\K$ are the shorthands for the time-translation, space-translation and boost generators,
respectively.  $[\P,\P]=l^{-2}\J$ implies
$[\P^+_i,\P^+_j]= - l^{-2} \eps_{ij}^{\ \ k}\J_k$,
etc. ($\epsilon_{123}=-\eps_{12}^{\ \ 3}=1, \ \ \eta_{ij}=-\delta_{ij}$.) The commutators
between $\J$s and the commutators between  $\J$ and ${\cal H}$, $\P$,  $\K$ are not
included in Table 1 because they have the
same form for different algebras.  Apart from the three classical geometrical algebras,
there are 10 more possible
algebras than those in BLL paper.
\begin{table}[h]
\tbl{All possible relativistic, geometrical and
non-relativistic kinematical algebras}
{\scriptsize
\begin{tabular}{cccccccc}
\hline Algebra & Symbol & Generator set\footnote{}
  &$[\cal H,\P]$&$[\cal H,\K]$&$[\P,\P]$&$[\K,\K]$&$[\P,\K]$\\
\hline \dS  &$\mathfrak{d}_+$ & $(H^+, \P^+_i, \K_i, \J_i)$ &
$\nu^2\K$&$\P$&$l^{-2}\J$&
$-c^{-2}\J$&$c^{-2}\cal H$\\
\AdS & $\mathfrak{d}_-$ &  $(H^-, \P^-_i, \K_i, \J_i)$ &$-\nu^2\K$
&$\P$&$-l^{-2}\J $
&$-c^{-2}\J$&$c^{-2}\cal H$\\
{\it Poincar\'e} &$\begin{array}{c}\mathfrak{p}\\
\mathfrak{p}_2\end{array}$ & $\begin{array}{c}(H, \P_i, \K_i, \J_i)\\
\{H', \P'_i, \K_i, \J_i\}\end{array}$ & 0 & $\P$ & 0 &$-c^{-2}\J$&$c^{-2}\cal H$\\
\hline {\it Riemann} & $\mathfrak{r}$&$(H^-, \P^+_i, \N_i,\J_i)$&
$-\nu^2\K$&\P&$l^{-2}\J$ &$
c^{-2}\J$&$-c^{-2}\cal H$\\
{\it Lobachevsky}&$\mathfrak{l}$&$(H^+, \P^-_i, \N_i,\J_i)$&
$\nu^2\K$ & \P & $-l^{-2}\J$
 &$c^{-2}\J$ &$-c^{-2}\cal H$\\
{\it Euclid}&$\begin{array}{c}\mathfrak{e}\\
\mathfrak{e}_2\end{array}$&$\begin{array}{c}(H, \P_i, \N_i, \J_i)\\
(-H', \P'_i, \N_i, \J_i)\end{array}$&0 &\P &0 &$c^{-2}\J$ &$-c^{-2}\cal H$\\
\hline
{\it Galilei}&$\begin{array}{c}\mathfrak{g}\\
\mathfrak{g}_2\end{array}$&$\begin{array}{c}(H, \P_i, \K^{\frak g}_i,\J_i)\\
(H', \P'_i, \K^{\frak c}_i,\J_i)\end{array}$&0 & \P  &  0 &   0 &   0   \\
{\it Carroll}&$\begin{array}{c}\mathfrak{c}\\
\mathfrak{c}_2\end{array}$ & $\begin{array}{c}(H, \P_i, \K^{\frak c}_i,\J_i )\\
(H', \P'_i, \K^{\frak g}_i,\J_i )\end{array}$ & 0& 0& 0& 0&$c^{-2}\cal H$\\
${NH}_+$  & $\begin{array}{c}\mathfrak{n_+}\\
\mathfrak{n}_{+2}\end{array}$ & $\begin{array}{c}(H^+, \P_i , \K^{\frak g}_i,\J_i )\\
(H^+, \P'_i, \K^{\frak c}_i,\J_i )\end{array}$&$\nu^2\K$ &\P & 0 & 0 &0 \\
${NH}_-$  &$\begin{array}{c}\mathfrak{n}_-\\
\mathfrak{n}_{-2}\end{array}$ & $\begin{array}{c}(H^-, \P_i, \K^{\frak g}_i,\J_i )\\
(-H^-, \P'_i, \K^{\frak c}_i,\J_i )\end{array}$ & $-\nu^2\K$ &\P &0 &0 &0 \\
{\it para-Galilei}&$\begin{array}{c}\mathfrak{g}'\\
\mathfrak{g}'_2\end{array}$ & $\begin{array}{c}(H', \P, \K^{\frak g}, \J_i)\\
(H, \P'_i, \K^{\frak c}_i,\J_i )\end{array}$ & $\nu^2\K$ & 0 & 0 & 0 &0 \\
$HN_+$\footnote{}&$\begin{array}{c}\mathfrak{h}_+\\
\mathfrak{h}_{+2}\end{array}$&$\begin{array}{c}(H, \P^+_i, \K^{\frak c}_i,\J_i )\\
(H', \P^+_i, \K^{\frak g}_i,\J_i )\end{array}$&$\nu^2\K$ &0 &$l^{-2}\J$&0 &$c^{-2}\cal H$\\
$HN_-$&$\begin{array}{c}\mathfrak{h_-}\\
\mathfrak{h}_{-2}\end{array}$&$\begin{array}{c}(H, \P^-_i, \K^{\frak c}_i,\J_i)\\
(-H', \P^-_i, \K^{\frak g}_i,\J_i )\end{array}$&$-\nu^2\K$ &0 &$-l^{-2}\J$&0 &$c^{-2}\cal H$\\
\hline {\it Static}&$\begin{array}{c}\mathfrak{s}\\
\mathfrak{s}_2\end{array}$&$\begin{array}{c}(H^{\frak s}, \P'_i,
\K^{\frak c}_i,\J_i ) \\
(H^{\frak s}, \P_i , \K^{\frak g}_i,\J_i )\end{array}$\footnote{}&0&0&0&0&0\\
\hline
\end{tabular}
}
\begin{tabnote}
$^{\rm b}$All commutators of
generators \omits{in (\ref{eq: JHPK})} have right dimensions expressed by
the universal constants $c, l$ or $\nu$.
$^{\rm c}$We named isomorphic to ISO(1,3) and para-Poincar\'e by the Hooke-Newton
and anti-Hooke-Newton algebras because of the relation between the generators of the two algebras
and the generators
of Newton-Hooke and anti-Newton-Hooke algebras.
$^{\rm d}$The generator
$H^{\frak{s}}$ is meaningful
only when the central extension is considered.
\end{tabnote}
\end{table}

In particular, both the set of generators $( H, {\bf P}_i, {\bf K}_i, {\bf J}_i)$ and
the set of generators $( H', {\bf P}'_i, {\bf K}_i, {\bf J}_i)$ satisfy the Poincar\'e
algebra
\be
&& [ {\bf P}_i, {\bf P}_j ] =0, \qquad  \ \   [ {\bf K}_i, {\bf K}_j ]
= - c^{-2}\eps_{ijk}{\bf J}_k, \  [ {\bf P}_i, {\bf K}_j ] = c^{-2}H, \qquad \nno \\
I:&& [ H, {\bf P}_i ] = 0,  \qquad \ \ \  [H, {\bf K}_i ] = {\bf P}, \qquad \qquad \quad
  [ {\bf J}_i, H ]=0,  \\
&& [ {\bf J}_i, {\bf P}_j ] = \eps_{ijk}{\bf P}_k, \ \, [ {\bf J}_i, {\bf K}_j ]
= \eps_{ijk}{\bf K}_k,  \qquad \ \ [ {\bf J}_i, {\bf J}_j ]= \eps_{ijk}{\bf J}_k,\nno
\ee
\be
&&[ {\bf P}'_i, {\bf P}'_j ] =0,
\qquad  \ \ [ {\bf K}_i, {\bf K}_j ] = - c^{-2}\eps_{ijk}{\bf J}_k,
\ [ {\bf P}'_i, {\bf K}_j ] = c^{-2}H',\qquad \nno \\
II:&&[ H', {\bf P}'_i ] = 0,   \qquad \ \  [H', {\bf K}_i ] = {\bf P}',
\qquad \qquad \ [ {\bf J}_i, H' ]=0,   \\
&&[ {\bf J}_i, {\bf P}'_j ] = \eps_{ijk}{\bf P}'_k,  \ \, [ {\bf J}_i, {\bf K}_j ]
= \eps_{ijk}{\bf K}_k, \qquad \ \; [ {\bf J}_i, {\bf J}_j ]= \eps_{ijk}{\bf J}_k,\nno
\ee
respectively.  The former set of generators are the generators of the
ordinary Poincar\'e transformations:
\be
{x'}^\mu = L^\mu_{\ \nu}x^\nu+ l a^\mu, \qquad L\in SO(1,3)
\ee
which can be realized by $5\times5$ matrix
\be  \label{PT}
\left(
  \begin{array}{cc}
    L & a \\
    0 & 1 \\
  \end{array}
\right).
\ee
The ordinary Poincar\'e transformation preserves the metric of the Minkowski
space-time.
The latter set of generators are the generators of transformations:
\be
{x'}^\mu = \d {L^\mu_{\ \nu}x^\nu}{1 +b_\mu x^\mu},
\ee
which can be expressed in terms of matrixes,
\be \label{2PT}
\left(
  \begin{array}{cc}
    L & 0 \\
    b^t & 1 \\
  \end{array}
\right). 
\ee
The set of all matrixes of type (\ref{2PT}) are the transpose of the set of all matrixes of
type (\ref{PT}) in
the Minkowski space-time.  Therefore, the set of all matrix (\ref{2PT}) also form
a Poincar\'e group.  It should be noted that the new Poincar\'e group
does not preserves the metric of the Minkowski
space-time.  Instead, it preserves the light cone at origin in the
Minkowski space-time\cite{GHWZ,GWZ}.

\section{New Geometries with Poincar\'e Symmetry}
Now that the new Poincar\'e group does not preserve the metric of
the Minkowski space-time, what geometry does the new Poincar\'e group
preserve?

It can be checked that $(M^\pm,\vect{g}^\pm,\vect{h}_\pm,\del^\pm)$ are invariant under
the new Poincar\'e
transformations, where $\vect{g}^\pm$ is a 4d type-(0,2) degenerate symmetric
tensor field
\be \label{g1}%
\vect{g}^\pm &=& g^\pm_{\mu\nu}dx^\mu \otimes dx^\nu
= \pm \frac {l^2} {(x \cdot x)^2}(\eta_{\mu \rho}\eta_{\nu \tau}
-\eta_{\mu\nu}\eta_{\rho \tau})x^\rho x^\tau dx^\mu dx^\nu,
\ee
$\vect{h}_\pm$ is a 4d type-(2,0) degenerate symmetric tensor field
\be
\label{gp2inv}
\vect{h}_\pm=h_\pm^{\mu\nu}\r_\mu \otimes \r_\nu =
l^{-4}(x\cdot x)x^\mu x^\nu\r_\mu \r_\nu,
\ee%
and $\del^\pm$ is a connection compatible to ${\vect g}^\pm$ and ${\vect h}_\pm$, i.e.
\be
\del^\pm_\la \, g_{\mu\nu}^\pm= \r_\la g^\pm_{\mu\nu}-\Gamma^\ka_{\pm\la\nu}g^\pm_{\mu\ka}
-\Gamma^\ka_{\pm\mu\la}g^\pm_{\ka\nu} =0
\ee
and
\be
\del^\pm_\la \,
{h}_\pm^{\mu\nu}= \r_\la {h}_\pm^{\mu\nu}+ \Gamma^\nu_{\pm\la\ka}{h}_\pm^{\mu\ka}
+\Gamma^\mu_{\pm\la\ka}{h}_\pm^{\ka\nu}=0,
\ee
respectively, with connection coefficients,
\be
\label{inv-connectn}%
 \Gamma^\mu_{\pm \nu\la} = - \d {(x_\nu
\dl^\mu_\la+\dl^\mu_\nu x_\la) }{x\cdot x}.
\ee
In the above equations,
\be
x \cdot x=\eta_{\mu\nu} x^\mu x^\nu \begin{cases} < 0 & \mbox{for upper sign},\\
                                                  > 0 & \mbox{for lower sign}.
                                                  \end{cases}
\ee
Clearly, $|g|=|h|=0$. The ranks of $\vect{g}$ and $\vect{h}$ are 3 and 1, respectively.
It can be shown that when and only when $\forall \vect{\xi} \in {\frak p}_2 \subset TM^\pm$,
\be
\begin{cases} \smallskip
{\cal L}_{\vect \xi} g^\pm_{\mu\nu}= g^\pm_{\mu\nu,\la}\vect{\xi}^\la+g^\pm_{\mu\la}\r_\nu
\vect{\xi}^\la
+g^\pm_{\la\nu}\r_\mu\vect{\xi}^\la=0, & \smallskip\\
{\cal L}_{\vect \xi} {h^\pm}^{\mu\nu}={h^\pm}^{\mu\nu}_{\ \ ,\la}
\vect{\xi}^\la -
{h^\pm}^{\mu\la}\r_\la\vect{\xi}^\nu
- {h^\pm}^{\la\nu}\r_\la\vect{\xi}^\mu=0, &\smallskip \\
[{\cal L}_{\vect \xi}, \del^\pm ] =0&
\end{cases}
\ee
are valid simultaneously.  By definition, the curvature is
\be
R^\mu_{\pm \nu \la \si}= l^{-2}(g^\pm_{\nu\la}\dl^\mu_\si-g^\pm_{\nu\si}\dl^\mu_\la)
\ee
and
\be\label{riccicurv}
R^\pm_{\mu \nu }= R^\la_{\pm \mu \nu \la}
= 3 l^{-2}g^\pm_{\mu\nu}.
\ee
\omits{Hence, the space-times are constant-curvature ones.}

In order to see the structures of the manifolds more transparently, we consider the
coordinate transformations,
\be
\label{Transf2dS}
\begin{array}{l}
x^0=l^2\rho^{-1} \sinh (\psi/l)  \\
x^1=l^2\rho^{-1} \cosh (\psi/l) \sin \th \cos\phi \\
x^2=l^2\rho^{-1} \cosh (\psi/l) \sin \th \sin \phi\\
x^3=l^2\rho^{-1} \cosh (\psi/l) \cos \th
\end{array}\quad \mbox{for }x\cdot x<0,
\ee
\be \label{Transf2AdS}
\begin{array}{l}
x^0=l^2\eta^{-1} \cosh (r/l)  \\
x^1=l^2\eta^{-1} \sinh (r/l) \sin \th \cos\phi \\
x^2=l^2\eta^{-1} \sinh (r/l) \sin \th \sin \phi\\
x^3=l^2\eta^{-1} \sinh (r/l) \cos \th
\end{array} \quad \mbox{for }x\cdot x>0,
\ee
respectively. Under the coordinate transformations, Eqs.(\ref{g1}), (\ref{gp2inv}), and
(\ref{inv-connectn}) become, respectively
\be \label{g3}
{\vect g}^\pm
=\begin{cases}d\psi^2 -l^2\cosh^2(\psi/l) d\Om_2^2 & {\rm for}\ x\cdot x <0\\
-d r^2 -l^2\sinh^2(r/l) d\Om_2^2& {\rm for}\ x\cdot
x>0,\end{cases}
\ee
\be\label{h1}
{\vect h}_\pm=\begin{cases}- \left (\d {\r }{\r \rho}\right)^2 & {\rm for}\ x\cdot x <0
\smallskip \\
 \left (\d {\r }{\r \rho}\right)^2 &{\rm for}\ x\cdot x <0,
\end{cases}
\ee
\be \label{connect+}\begin{array}{l}
{\bar \Gamma}^\psi_{+\th\th}=l\sinh(\psi/l) \cosh(\psi/l),\quad
{\bar \Gamma}^\psi_{+\phi\phi}=
{\bar \Gamma}^\psi_{+\th\th}\sin^2\th \\
{\bar \Gamma}^\th_{+\th\psi}={\bar \Gamma}^\th_{+\psi\th}={\bar \Gamma}^\phi_{+\phi\psi}
={\bar \Gamma}^\phi_{+\psi\phi}
=l^{-1} \tanh(\psi/l) \\
{\bar \Gamma}^\th_{+\phi\phi}=- \sin \th \cos \th , \quad
{\bar \Gamma}^\phi_{+\th\phi}={\bar \Gamma}^\phi_{+\phi\th}= \cot \th \\
{\bar \Gamma}^\rho_{+\al\beta} = -l^{-2}\rho g_{\al\beta}, \qquad
\mbox{others vanish},\end{array}\quad {\rm for} \ x\cdot x <0,
\ee
where $(\bar x^\al; \bar x^3)
=(\psi, \th, \phi; \rho)$, and
\be\label{connect-}\begin{array}{l}
{\bar \Gamma}^\eta_{-ij} = -l^{-2}\eta g_{ij}\\
{\bar \Gamma}^r_{-\th\th} =-l\sinh(r/l) \cosh(r/l), \quad
{\bar \Gamma}^r_{-\phi\phi} ={\bar \Gamma}^r_{-\th\th}\sin^2\th   \\
{\bar \Gamma}^\th_{-r\th} ={\bar \Gamma}^\th_{-\th r} ={\bar \Gamma}^\phi_{-r\phi}
={\bar \Gamma}^\phi_{-\phi r}
=l^{-1}{\rm ctanh}(r/l) \\
{\bar \Gamma}^\th_{-\phi\phi} =-\sin\th\cos\th, \quad
{\bar \Gamma}^\phi_{-\th\phi}={\bar \Gamma}^\phi_{-\phi\th}=\cot\th \\
\mbox{others vanish,}
\end{array}  \quad {\rm for}\ x\cdot x >0,%
\ee
where $(\bar x^0;\bar x^i)=(\eta; r, \th,
\phi)$.
The Ricci curvature (\ref{riccicurv}) read
\be %
R_{\mu\nu}^\pm
= \begin{cases} 3l^{-2}\,{\rm diag}(1, -\cosh^2(\psi/l),-\cosh^2(\psi/l)
\sin^2\th, 0) & x\cdot x<0, \\
-3l^2\,{\rm diag}(0, -1, -\sinh^2(r/l), -\sinh^2(r/l) \sin^2\th)&
x\cdot x >0.
\end{cases}
\ee
They show that the
space-times are the homogeneous spaces and that
\be
M^+=ISO(1,3)/ISO(1,2) =\mathbb{R}\times dS_3 \qquad {\rm for} \qquad x\cdot x<0,
\ee
and
\be
M^-=ISO(1,3)/ISO(3)=\mathbb{R}\times H_3  \qquad {\rm for} \qquad x\cdot x>0.
\ee
It is obvious that on $M^+$ there is no spatial SO(3) isotropy at each point
though there exists the algebraic SO(3) isotropy, but on $M^-$ there exists
the spatial SO(3) isotropy at each point.

Further studies show that $(M^-,\vect{g}^-,\vect{h}_-,\del^-)$ satisfies
the three basic assumptions in the Theorem in Ref. \refcite{BLL}.  The kinematics
on it will have better behaviors than on $(M^+,\vect{g}^+,\vect{h}_+,\del^+)$.  Therefore,
we shall study the kinematics briefly.

\section{Motion of a Free Particle on $(M^-,\vect{g}^-,\vect{h}_-,\del^-)$}
\omits{Since $(M^-,\vect{g}^-,\vect{h}_-,\del^-)$ possesses the spatial SO(3) isotropy,
 kinematics on it.} The motion for a
free particle is still supposed to be determined by the
geodesic equation%
\be
\d {d^2 x^\mu}{d\la^2}+\Gamma^\mu_{-
\nu\la}\d{dx^\nu}{d\la}\d{dx^\la}{d\la}=0,
\ee
as usual.  It gives rise to the `uniform rectilinear' motion%
\be
\label{gim}
 x^i= a^i x^0 + lb^i
\ee
if $x^0$ and $x^i$ are regarded as the `temporal' and `spatial' coordinates,
respectively, where  $a^i$ and $b^i$ are two dimensionless constants.  The
result is consistent with Eq.(\ref{urm}) and Eq.(\ref{a0}), which are the
start points of our work.

The `uniform rectilinear' motion (\ref{gim}) can also be obtained from the
Lagrangian
\be \label{Lagrangian}
L= \frac {mlc} {x \cdot x}\sqrt{(\eta_{\mu\nu}\eta_{\eta \tau}
-\eta_{\mu \eta}\eta_{\nu \tau}
)x^\eta x^\tau \dot x^\mu \dot x^\nu}.
\ee
The Euler-Lagrangian equation is equivalent to
\be
[(x\cdot x)(\dot x\cdot \dot x)-(x\cdot \dot x)^2 ]
\ddot x_\ka\ +(\dot x\cdot \ddot x)[(x\cdot \dot x) x_\ka -(x\cdot x)
\dot x_\ka] \nno\\
+(x\cdot \ddot x)[(x\cdot \dot x)\dot x_\ka
-(\dot x\cdot \dot x)x_\ka]=0. \qquad \qquad \qquad\nno
\ee
The nonzero determinant of its coefficients for $\ddot x$
requires $\ddot x_\ka= 0.$

It should be noted that the coordinate system $x^0/c$ and $x^i$ are not the
intrinsic coordinates of the time and space, respectively.  Therefore, Eq.(\ref{gim})
cannot be interpreted as the uniform rectilinear motion or inertial motion in the
space-time $(M^-,\vect{g}^-,\vect{h}_-,\del^-)$ in the usual sense.

\section{Summary}

There are 24 different possible kinematical groups, including geometrical
ones and static one.  Each has 10 parameters.  Among the 24 possible kinematical
groups, there exists a new Poincar\'e symmetry in addition to the ordinary
Poincar\'e symmetry.

The new Poincar\'e symmetry does not preserve the metric of the Minkowski space-time.
Instead, it preserves the degenerate geometries $(M^\pm, \vect{g}^\pm,
\vect{h}_\pm, \del^\pm)$ presented in the talk.  The degenerate geometries and
their topology are dramatically different from those of the Minkowski space-time.
But, they are still homogeneous spaces.  The physical
applications of the degenerate space-times need to be explored.

From the study on the degenerate geometries with Poincar\'e symmetry, we can see
that algebraic SO(3) isotropy does not always imply the geometrical space isotropy.
Whether a kinematics possesses the space isotropy or space-time isotropy should be
determined by the underlying geometry.

The Lagrangian for a free particle can be defined on the new geometry.  In the
coordinate system $x^\mu$, the motion takes the form of the uniform rectilinear motion.
Unfortunately, the coordinate system $x^i$ and $x^0/c$ do not the intrinsic coordinates
of the space and time.  Therefore, `the uniform rectilinear motion'
is not in the usual sense.

\section*{Acknowledgments}

I am grateful to H.-Y. Guo, Y. Tian,  H.-t. Wu, X.-N. Wu, Z. Xu,
and B. Zhou for the cooperation in the works related to the talk.
I would like to thank Z.-N. Hu, W. T. Ni, J. Xu, and H.-X. Yang for
helpful discussion. The work is supported in part by NSFC under Grant Nos. 10775140,
10975141, and KIFCAS (KJCX3-SYW-S03).

\end{document}